\documentclass{sig-alternate}

\usepackage[pass]{geometry}

\newfont{\mycrnotice}{ptmr8t at 7pt}
\newfont{\myconfname}{ptmri8t at 7pt}

\let\crnotice\mycrnotice%
\let\confname\myconfname%

\permission{Permission to make digital or hard copies of all or part of this work for personal or classroom use is granted without fee provided that copies are not made or distributed for profit or commercial advantage and that copies bear this notice and the full citation on the first page. Copyrights for components of this work owned by others than ACM must be honored. Abstracting with credit is permitted. To copy otherwise, or republish, to post on servers or to redistribute to lists, requires prior specific permission and/or a fee. Request permissions from permissions@acm.org.}
\conferenceinfo{PE-WASUN'14,}{September 21--26, 2014, Montreal, QC, Canada.}
\copyrightetc{Copyright 2014 ACM \the\acmcopyr}
\crdata{978-1-4503-3025-1/14/09\ ...\$15.00.\\
http://dx.doi.org/10.1145/2653481.2653484}
\usepackage{graphicx}
\usepackage[font={footnotesize}]{caption}
\usepackage{amsmath}

\usepackage{hyphenat}
\usepackage{ragged2e}

\usepackage{amsmath}
\usepackage{url}
\usepackage[numbers]{natbib}
\begin{document}


\title{How to Choose the Relevant MAC Protocol for Wireless Smart Parking Urban Networks?}
%
%
%
%
%

%
\author{
\and
Trista Lin, Herv{\'e} Rivano\\
       \affaddr{INRIA, Universit{\'e} de Lyon}\\
       \affaddr{INSA-Lyon, CITI-Inria}\\
       \affaddr{F-69621, Villeurbanne, France}\\
       \email{\{trista.lin, herve.rivano\}@inria.fr}
\and
Fr{\'e}d{\'e}ric Le Mou{\"e}l\\
       \begin{tabular}[t]{@{}c@{}}
       \affaddr{Universit{\'e} de Lyon}\\
       \affaddr{INSA-Lyon, CITI-Inria}\\
       \affaddr{F-69621 Villeurbanne, France}\\
       \end{tabular}\nobreak\qquad
       \begin{tabular}[t]{@{}c@{}} 
       \affaddr{Shanghai JianTong University} \\
       \affaddr{No. 800 DongChuan Rd.} \\
       \affaddr{Shanghai, China} \\
       \end{tabular} \\
       \email{frederic.le-mouel@insa-lyon.fr}
}

\maketitle
\begin{abstract}
Parking sensor network is rapidly deploying around the world and is regarded as one of the first implemented urban services in smart cities. To provide the best network performance, the MAC protocol shall be adaptive enough in order to satisfy the traffic intensity and variation of parking sensors. In this paper, we study the heavy-tailed parking and vacant time models from SmartSantander, and then we apply the traffic model in the simulation with four different kinds of MAC protocols, that is, contention-based, schedule-based and two hybrid versions of them. The result shows that the packet interarrival time is no longer heavy-tailed while collecting a group of parking sensors, and then choosing an appropriate MAC protocol highly depends on the network configuration. Also, the information delay is bounded by traffic and MAC parameters which are important criteria while the timely message is required. 
\end{abstract}




\section{Introduction}

As the problem of parking search increases, networked parking sensor device is being installed everywhere to detect the availability of parking spaces. These parking sensors form a wireless network (WSN) and is called parking sensor network (PSN). PSN has the following characteristics: First, parking sensors are stationary, in-ground and scattered with a minimum adjacent distance. Second, the network topology is linear and limited by urban street layout. Third, the sensing area of each parking sensor does not have any intersection because of the lack of multiple detection. Fourth, packet generation rate depends on the vehicle's arrival and departure. Fifth, the available parking information is the data of real-time parking service. Hence, we can say that PSN is a specialized form of WSN and also inherits its energy and delay constraints. Based on these, we see the importance of the device lifetime, considering their maintenance and latency while providing real-time service to urban citizens. MAC protocols, dealing with the network resource allocation and the ON/OFF radio states, mainly affect the delay time and consumed energy. Therefore, what are the criteria to choose a good MAC protocol in order to pursue a high performance of WSN? Before studying the network performance, the most important is to understand what kind of network traffic to deal with. It is shown that the network traffic intensity in PSN is very variable and heavy-tailed during a day. We took the parameters from literature and then applied into our simulation to see how it affects the packet interarrival time. If a mixed periodic and event-drive application is used on sensors, the traffic interarrival time can be also reshaped. Next, with these traffic parameters, what kind of MAC protocol can adapt better with such kind of network traffic and topology? We studied the impact on the criteria of two main off-the-shelf MAC protocols and their hybrid versions. Our contribution are summarized as below:

\begin{itemize}
	\item Based on the well-known model to describe vehicle's arrival and departure from literature, we show that while collecting a group of nodes, the packet and vehicle interarrival times are no longer heavy-tailed. 
  \item We introduce a network architecture to collect the data of smart parking. Sensors are connected to routers and these routers are interconnected. Devices are classified as either a full function device or a reduced function device in order to achieve the best energy-efficiency for urban infrastructure.  
	\item A performance comparison of four different MAC protocols is shown by evaluating their information delay and energy consumption in parking sensor networks.
\end{itemize}

The remainder of this paper is structured as follows. In the Section~\ref{sec:relatedworks}, we give a review on three groups of different MAC protocols. In Section~\ref{sec:networktrafficmodel} we introduce the network traffic model and the sun of two Weibull variates. In Section~\ref{sec:mac_protocol} gives an introduction for the MAC protocols we evaluate. In Section~\ref{sec:experience} we construct the urban environment and then perform the simulations. Finally, we summarize our works in Section~\ref{sec:conclusion}.

\section{Related Work}
\label{sec:relatedworks}

Many MAC protocols have been proposed for urban wireless sensor network application. Here, we sort them as the following three groups:

\begin{itemize}
	\item \textsc{Contention-based} protocols are much widely studied in WSN and generally based on or similar to CSMA /CA. When one node has a packet to send, it will have to struggle with the other competitors to get permission to use the medium. The winner selection is somehow randomized. The synchronous MAC protocols are generally duty-cycled and require time-synchronized, such as S-MAC, T-MAC, Conti\cite{ctcontention2005}, SIFT\cite{jamieson2006sift} and so forth. The state-of-the-art synchronization method is to be done through hardware or message exchange, and then a piggybacked acknowledge can be used to solve the clock shifting effect\cite{6488838}. The asynchronous versions use \textit{low-power listening}(LPL) or its preamble-shortened approach to match up the transmission period between transmitter and receiver end, such as B-MAC, WiseMac, X-MAC, SCP-MAC\cite{Ye:2006:UDC:1182807.1182839}, Ri-MAC\cite{Sun:2008:RRA:1460412.1460414} and so on. Among them, \cite{4622710,6214027,thesisdequentin} compared the power dissipation between asynchronous and synchronous contention-based protocol. In which, LPL method is interesting in very low traffic intensity (less than one packet per day) or dynamically changing topologies. Otherwise, synchronous protocol outperforms asynchron\-ous one. The drawback of such a protocol is the packet collision caused by increasing network density and hidden terminal. 
	
	\item \textsc{Schedule-based} protocols are generally centralized and suitable for static topologies. Assigned nodes play the master role to allocate slotted network resource to their slaves. The mechanism is generally based on TDMA or CDMA. The clock of each node must be time-synchronized. Scheduled slots can be fixed or on-demand. Some noted protocols are like \textsc{Drand}\cite{Rhee:2006:DDR:1132905.1132927}, Leach\cite{926982}, Trama\cite{Rajendran:2003:ECM:958491.958513} and TSMP\cite{Pister08tsmp:time}. They use TDMA as the baseline \textsc{Mac} scheme, and then take CSMA, Aloha or CDMA for improving these join/leave/synch-ronize messages. The drawbacks are firstly not easy to adapt to the dynamics of network, and secondly the slower response of the centralized control while adapting the schedule to the traffic variation. 
		
	\item \textsc{Hybrid contention- \& schedule-based} combine the advantages of both protocols in order to reach the best performance. \textsc{Z-mac}\cite{4453818} behaves like CSMA under low competition and behaves like TDMA under high competition. Funneling-MAC\cite{Ahn:2006:FLS:1182807.1182837} uses CSMA as the baseline, and changes to TDMA while receiving on-demand beaconing from the sink, that is to say, nodes close to the sink performs \textsc{Tdma}. \textsc{Funneling-mac} works in the application of data collection. iQueue-mac\cite{6644967} runs in CSMA in light load and then uses queue-length piggybacking as accurate load information to ask for additional variable TDMA time slots if needed. 	 
\end{itemize}

From literature, we then study the traffic model in PSNs and test them by diverse MAC protocols from the three mainstream categories. 


\section{Network traffic formulation}
\label{sec:networktrafficmodel}

\subsection{Vehicle event detection}

\begin{figure}[!t]
\centering
\includegraphics[width=\linewidth]{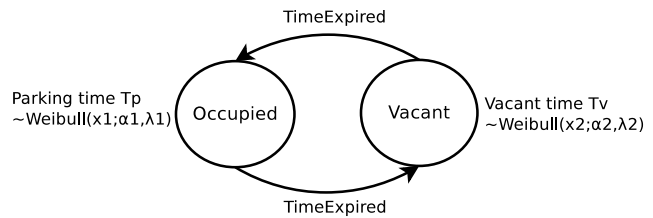}
\caption{Occupied/vacant state transition diagram}
\label{fig:weibull-sensor-state}
\end{figure}

\begin{figure}[!t]
\centering
\includegraphics[height=\linewidth, angle=-90]{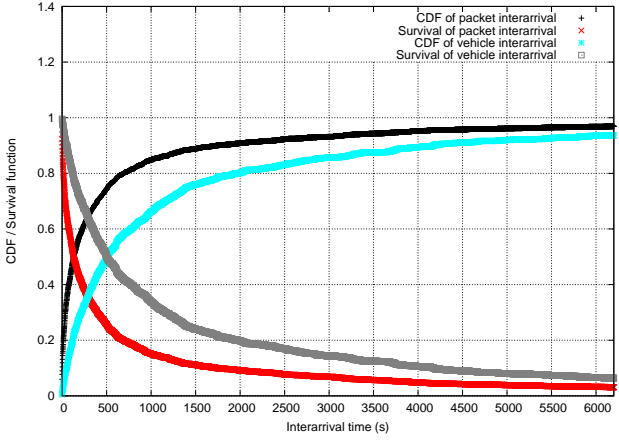}
\caption{The CDF and survival function of vehicle and packet interarrival times of one in-ground parking sensor}
\label{fig:packet-vehicle-interarrival-time-per-node}
\end{figure}

In PSNs, the main observed event are vehicles' arrivals and departures. For any parking sensor, each vehicle arrival accompanies exact by one departure prior to next arrival. To model it, we first look at the event occurrence sequence on one parking sensor. We suppose that each parking sensor is precise enough and provides merely two states, namely occupied and vacant, shown in figure~\ref{fig:weibull-sensor-state}. The occupied time from vehicle's arrival to departure is so-called parking time $T_{p}$, conversely, the vacant time is available time $T_{v}$. During which, each sensor detects the vehicle's presence or absence. Both $T_{p}$ and $T_{v}$ shall be described by a fitting distribution in order to approximate their randomness. From Vlahogianni's report\cite{SmartSantanderReport}, the massive real-time parking availability data, obtained by 4 on-street PSNs in Santander, shows that the occupied duration is best described by a Weibull distribution. Besides, the findings in \cite{SmartSantanderReport} show that the duration of free parking space follows a Weibull distribution as well. By assuming that $T_{p}$ and $T_{v}$ are both Weibull distributed, the PDF of $T_{p}$ and $T_{v}$ is given by: 

\vspace{-2mm}
\begin{itemize}
  \item $f_{X_1}(x_1) = \frac{\alpha_1}{\lambda_1^{\alpha_1}}x_1^{\alpha_1-1}\mathrm{e}^{-(\frac{x_1}{\lambda_1})^{\alpha_1}}$ stands for the probability of choosing a parking time $X_1$. 
  \item $f_{X_2}(x_2) = \frac{\alpha_2}{\lambda_2^{\alpha_2}}x_2^{\alpha_2-1}\mathrm{e}^{-(\frac{x_2}{\lambda_2})^{\alpha_2}}$ stands for the probability of choosing a vacant time $X_2$.
\end{itemize}

where $0<\alpha_i<1$ is a shape parameter and $\lambda_i$ is a scale parameter, for $i=1,2$. Let $X$ be the sum of two i.i.d. Weibull variates $X_i$, i.e., $X=X_1+X_2$. The approximate PDF and CDF of $X$ proposed in \cite{1665128} are given by:

\begin{equation}
f_{X}(x) = \frac{\alpha}{(\lambda^{\alpha)^\mu}} \frac{\mu^{\mu}}{\Gamma(\mu)}  x^{\alpha\mu -1}\mathrm{e}^{-\mu(\frac{x}{\lambda})^{\alpha}}
\end{equation}

\begin{equation}
F_{X}(x) = 1 - \frac{\Gamma(\mu, \mu (\frac{x}{\lambda})^{\alpha})}{\Gamma(\mu)}
\end{equation}

where $\alpha$ is a shape parameter, $\lambda$ is a scale parameter $\ni E[X^\alpha] = \lambda^\alpha$ and $\mu = E^{2}[X^{\alpha}]/Var(X^{\alpha}) > 0$.
$\Gamma(\cdot)$ is the gamma function and $\Gamma(\cdot, \cdot)$ the incomplete gamma function. If we take $\alpha_1 = 0.4$, $\lambda_1=3600(s)$, $\alpha_2 = 0.7$ and $\lambda_2 = 900(s)$ referring to \cite{SmartSantanderReport}, and then the vehicle interarrival time of one parking place is obtained as $T_p + T_v = X_1 + X_2 = X$ in figure ~\ref{fig:packet-vehicle-interarrival-time-per-node}. From the survival function of vehicle interarrival time, we see that $X$ is also heavy-tailed, i.e., $0<\alpha<1$. $\alpha$, $\mu$ and $\lambda$ can be obtained by equations (\ref{eq:getXparameter1})--(\ref{eq:getXparameter3}) written in \cite{1665128}.  

\begin{equation}
\frac{\Gamma^2(\mu + \frac{1}{\alpha})}{\Gamma(\mu)\Gamma(\mu + \frac{2}{\alpha}) - \Gamma^2(\mu + \frac{2}{\alpha})} = \frac{E^2[R]}{E[R^2]-E^2[R]}
\label{eq:getXparameter1}
\end{equation}

\begin{equation}
\frac{\Gamma^2(\mu + \frac{2}{\alpha})}{\Gamma(\mu)\Gamma(\mu + \frac{4}{\alpha}) - \Gamma^2(\mu + \frac{2}{\alpha})} = \frac{E^2[R^2]}{E[R^4]-E^2[R^2]}
\label{eq:getXparameter2}
\end{equation}

\begin{equation}
\lambda = \frac{\mu^{\frac{1}{\alpha}} \Gamma(\mu) E[R]}{\Gamma(\mu + \frac{1}{\alpha})}
\label{eq:getXparameter3}
\end{equation}

where $E[X^n]$ is the $n$$^{th}$ moment of $X$ and given by $E[X^{n}] = \sum\nolimits_{n_1=0}^{n}\sum\nolimits_{n_2=0}^{n_1} {n \choose n_1} {n_1 \choose n_2} E[X^{n-n_1}_1] E[X^{n_1-n_2}_2]$. $X_i$ is Weibull distributed so that its $n$$^{th}$ moment is $E[X^{n}_{i}] = \lambda_{i}^{n} \Gamma(1+\frac{n}{\alpha_{i}})$.

The count model based on Weibull interarrival times is studied in \cite{EricBradlow}. 
The probability of $k$ arriving vehicles in a given interval is given as below:

\begin{equation}
P(N(t)=k)=\sum_{j=k}^{\infty}\frac{(-1)^{j+k}(\frac{t}{\lambda})^{\alpha j} \Delta_{j}^{k}}{\Gamma(\alpha j+1)} \;\; k = 0,1,\cdots
\label{eq:prob_of_k_vehicle_arriving}
\end{equation}

where $\Delta_{j}^{k+1} = \sum\nolimits_{m=k}^{j-1} \Delta_{m}^{k} \frac{\Gamma(\alpha j - \alpha m +1)}{\Gamma(j - m +1)}$ for $k=0,1, 2,\cdots$ and $j=k+1,k+2,k+3,\cdots$. $\Delta_{j}^{0}=\frac{\Gamma(\alpha j+1)}{\Gamma(j+1)} \; j=0,1,2,\cdots$. In a business area or on weekdays, the average free time can be shorter according to the area hourely activities or parking policy. Its impact to network traffic will have to be considered as well. 

\begin{figure}[!t]
\centering
\includegraphics[height=\linewidth, angle=-90]{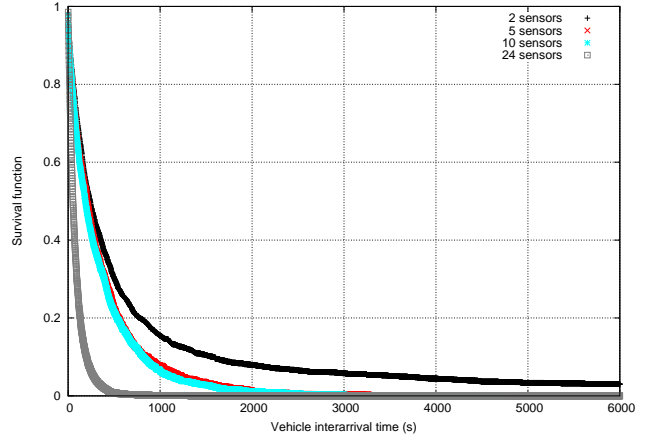}
\caption{The survival function of vehicle interarrival times of 2, 5, 10 and 24 in-ground parking sensors}
\label{fig:vehicle-interval-survival-2-5-10-24-nodes}
\end{figure}
\begin{figure}[!t]
\centering
\includegraphics[height=\linewidth, angle=-90]{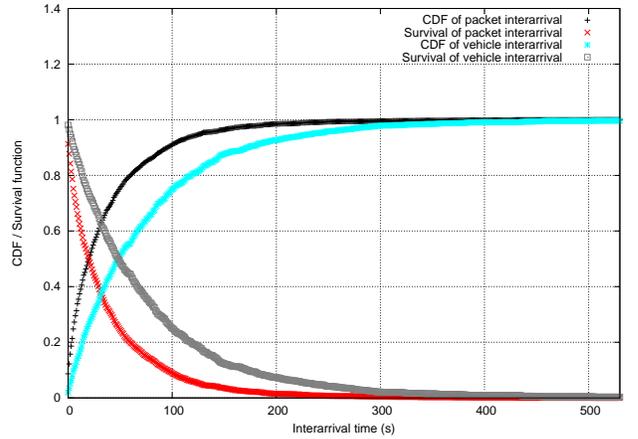}
\caption{The CDF and survival function of vehicle and packet interarrival times of 24 in-ground parking sensors}
\label{fig:packet-vehicle-interarrival-time-24-nodes}
\end{figure}

\subsection{Event-Driven traffic}
If the parking sensor sends a packet while changing the sensing state, the packet interarrival time $Y$ is confined to the vehicle's interarrival time and will be also heavy-tailed with the shape and scale parameters $A$ and $\Lambda$, where $A = \alpha$ and $\Lambda = 0.294\lambda$ in figure~\ref{fig:packet-vehicle-interarrival-time-per-node}. Next, we look at a group of parking sensors and evaluate the vehicle interarrival time for a parking area. Here, the vehicle interarrival time stands for the interval of any two consecutive vehicle arrivals of different parking spaces in this area. With the increasingly network size, the vehicle interrarrival time is approximating to exponential distribution, shown in figure~\ref{fig:vehicle-interval-survival-2-5-10-24-nodes}, i.e., $\alpha \rightarrow 1$. In which, the relationship between vehicle and packet interarrival time varies from $\Lambda = 0.25\lambda$ (2 sensors) to $0.225\lambda$ (24 sensors). Figure~\ref{fig:packet-vehicle-interarrival-time-24-nodes} shows that the packet and vehicle interarrival times of a 24-parking-sensor network follow the exponential distribution (the Weibull's shape parameter is equal to 1). 

\begin{figure}[!t]
\centering
\includegraphics[height=\linewidth, angle=-90]{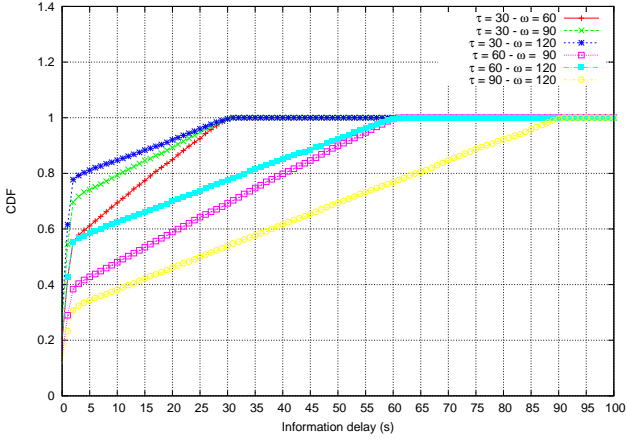}
\caption{Periodic and Event-driven mixed application}
\label{fig:periodic-trigger-mixed-app}
\end{figure}

Periodic application is only affected by the periodic time interval $\omega$ and widely used in WSN for periodic collection of energy status. By assuming that the sensory information is small enough to merge into the packet with energy status information, we define a time parameter $\tau$ as the threshold to decide when the information is going to be sent. When the sensing state changes at the time point $t_{now}$, it will check the delivery time of next periodic packet $t_{next}$. If $|t_{now} - t_{next}| < \tau$, the occupancy information will be put into the periodic packet to reduce the transmission frequency. Otherwise, this information will be canned into a packet and sent immediately. This way, all the parking information shall be delivered within $\tau$ seconds. In PSNs, we evaluate the information delay to estimate if the protocol is efficient enough. Information delay is the required time to know a changed occupancy status of a parking sensor. Figrue~\ref{fig:periodic-trigger-mixed-app} shows the CDF of information delay while applying hybrid periodic and event-driven application. For the generated information $\ni |t_{now} - t_{next}| < \tau$, the packets will be sent within $\tau$ seconds uniformly. Otherwise, the packets will be sent right now and will arrive at the destination in $1^{st}$ duty cycle ($T_{cycle}$) since the network traffic is not that heavy. The probability of packet arriving in $1^{st}$ duty cycle is given as $Pr\{I_i| t_{next} - t_{now}(I_i) >= \tau\} = \frac{\omega - \tau}{\omega} + \frac{T_{cycle}}{\omega}$, and the probability of packet arriving between $1^{st}$ duty cycle and $\tau$ is $Pr\{I_i| t_{next} - t_{now}(I_i) < \tau\} = \frac{\tau - T_{cycle}}{\omega}$, where $I_i$ is the $i^{th}$ sensed information. Hence, the smaller $\frac{\tau}{\omega}$ is, the shorter information delay we expect. 

\section{MAC protocols}
\label{sec:mac_protocol}

\begin{figure}[ht]
\renewcommand
\footnotesize
\centering
\begin{tabular}{|c|c|c|c|c|c|c|c|c|c|c|c|c|c|c|c|c|c|}
\multicolumn{18}{l}{CSMA} \\ \hline
\multicolumn{5}{|c|}{SYN}&&&\multicolumn{4}{c|}{$\cdots\cdots$}&&&&&\multicolumn{3}{c|}{\cellcolor{gray}} \\ \hline
\multicolumn{5}{c}{}&\multicolumn{10}{|c|}{$\longleftarrow$ CSMA slots $\longrightarrow$} & \multicolumn{3}{c|}{$T_{GTS}$} \\
\multicolumn{15}{l}{} & \multicolumn{3}{c}{$+T_{inactive}$} \\
\multicolumn{18}{l}{TDMA} \\ \hline
\multicolumn{5}{|c|}{SYN+Signaling} &&&&&\multicolumn{2}{c|}{$\cdots$}&&&&&\multicolumn{3}{c|}{\cellcolor{gray}}  \\ \hline
\multicolumn{5}{c}{}  & \multicolumn{10}{|c|}{$\longleftarrow$ TDMA slots $\longrightarrow$} & \multicolumn{3}{c|}{$T_{GTS}$} \\
\multicolumn{15}{l}{} & \multicolumn{3}{c}{$+T_{inactive}$} \\
\multicolumn{18}{l}{Funnelling-MAC} \\ \hline
\multicolumn{5}{|c|}{SYN+Signaling} &&&\multicolumn{2}{c|}{$\cdots$}&&&&\multicolumn{2}{c|}{$\cdots$}&& \multicolumn{3}{c|}{\cellcolor{gray}}\\ \hline
\multicolumn{5}{c}{} & \multicolumn{5}{|c|}{$\leftarrow$CSMA$\rightarrow$} & \multicolumn{5}{c|}{$\leftarrow$TDMA$\rightarrow$}& \multicolumn{3}{c|}{$T_{GTS}$} \\
\multicolumn{5}{c}{} & \multicolumn{5}{c}{slots}                           & \multicolumn{5}{c}{slots}              & \multicolumn{3}{c}{$+T_{inactive}$} \\
\multicolumn{18}{l}{i-Queue} \\ \hline
\multicolumn{5}{|c|}{SYN}&&\multicolumn{4}{c|}{$\cdots\cdots$}&&&\multicolumn{2}{c|}{$\cdots$}&&\multicolumn{3}{c|}{\cellcolor{gray}} \\ \hline
\multicolumn{5}{c}{}&\multicolumn{6}{|c|}{$\leftarrow$ CSMA $\rightarrow$} & \multicolumn{4}{c|}{vTDMA}& \multicolumn{3}{c|}{$T_{GTS}$}\\
\multicolumn{5}{c}{}&\multicolumn{6}{c}{slots} & \multicolumn{4}{c}{slots}& \multicolumn{3}{c}{$+T_{inactive}$}\\
\multicolumn{18}{l}{*SYN: synchronization} \\
\multicolumn{18}{l}{*GTS: Guarantee time slot} \\
\end{tabular}
\caption{Slot allocation methods of CSMA, TDMA, Funnelling-MAC and i-Queue MAC protocols}
\label{tab:MacProtocol}
\end{figure}

Considering that WSN is often bandwidth-limited, the only medium resource is time divisions in a single-channel scenario. From literature, we found that choosing a MAC protocol has been the subject of much controversy in urban sensor networks. We selected four different types of MAC protocols in order to evaluate the network performance. They are contention-based, schedule-based and hybrid from both of them respectively. As previously mentioned, these MAC protocols are all duty-cycled, time-slotted and run in a time-synchronized environment. Our work is to find the most appropriate MAC protocol to deal with the data transmission in the CAP(contention access period) of IEEE 802.15.4 MAC standard\cite{1395158}. For a better comparison in CAP, these four MAC protocols apply the same amount of time slots and the hybrid versions will adjust the percentage of TDMA or CSMA slots in order to combine the advantages of TDMA and CSMA in one protocol. $T_{cycle}= (n_{csma} + n_{tdma}) * s_D + T_{GTS} + T_{inactive}$ for $n$ is the slot number.

First, CSMA is the default version of CAP in the MAC layer of IEEE 802.15.4. CSMA is the typical contention-based protocol and widely applied in practice. The principle is that the node gets its partition of network resource when it asks for. If more than two nodes declare their demands, one competition will happen to decide who is the current transmitter. The transmitter candidates choose a backoff time for sending the reservation beacon and then listen to the medium before the backoff time expires. If these candidates hear any beacon in the meantime, they will turn off the radio and wait for the next time slot. 

TDMA is based on a preassigned schedule managed by the network coordinator. If a node has no packet to send in its item, the other still cannot seize this occasion to send their packets out. TDMA can achieve a very little packet collision rate but is inflexible for its central schedule control. A signaling process is always needed when the network topology changes. 

Funnelling-MAC \cite{Ahn:2006:FLS:1182807.1182837} is a hybrid version of TDMA and CSMA. All devices run CSMA by default. The devices, which are $N$-hop away from the gateway, will change to TDMA and then ask for a time slot to the gateway. When $N=1$, Funnelling-MAC works nearly like TDMA in one-hop scenario. Funnelling-MAC is equal to TDMA when $N=\#{\mbox{Max Hop}}$. 

i-Queue\cite{6644967} sets an one-byte queue indicator in the MAC header, so that the gateway will be informed of the additional demands from data packets and reply with some additional time slots from vTDMA (variable TDMA) slots. If the gateway receives no request of vTDMA slots, the duration of vTDMA slot will become inactive and useless for sensor nodes. Since the packet interarrival time is scattered in PSNs, i-Queue MAC behaves nearly like CSMA. Here, we assume that only the gateway can issue an additional vTDMA slots to the nodes one-hop far away.

\section{Evaluation}
\label{sec:experience}
\begin{figure}[!t]
\centering
\includegraphics[width=\linewidth]{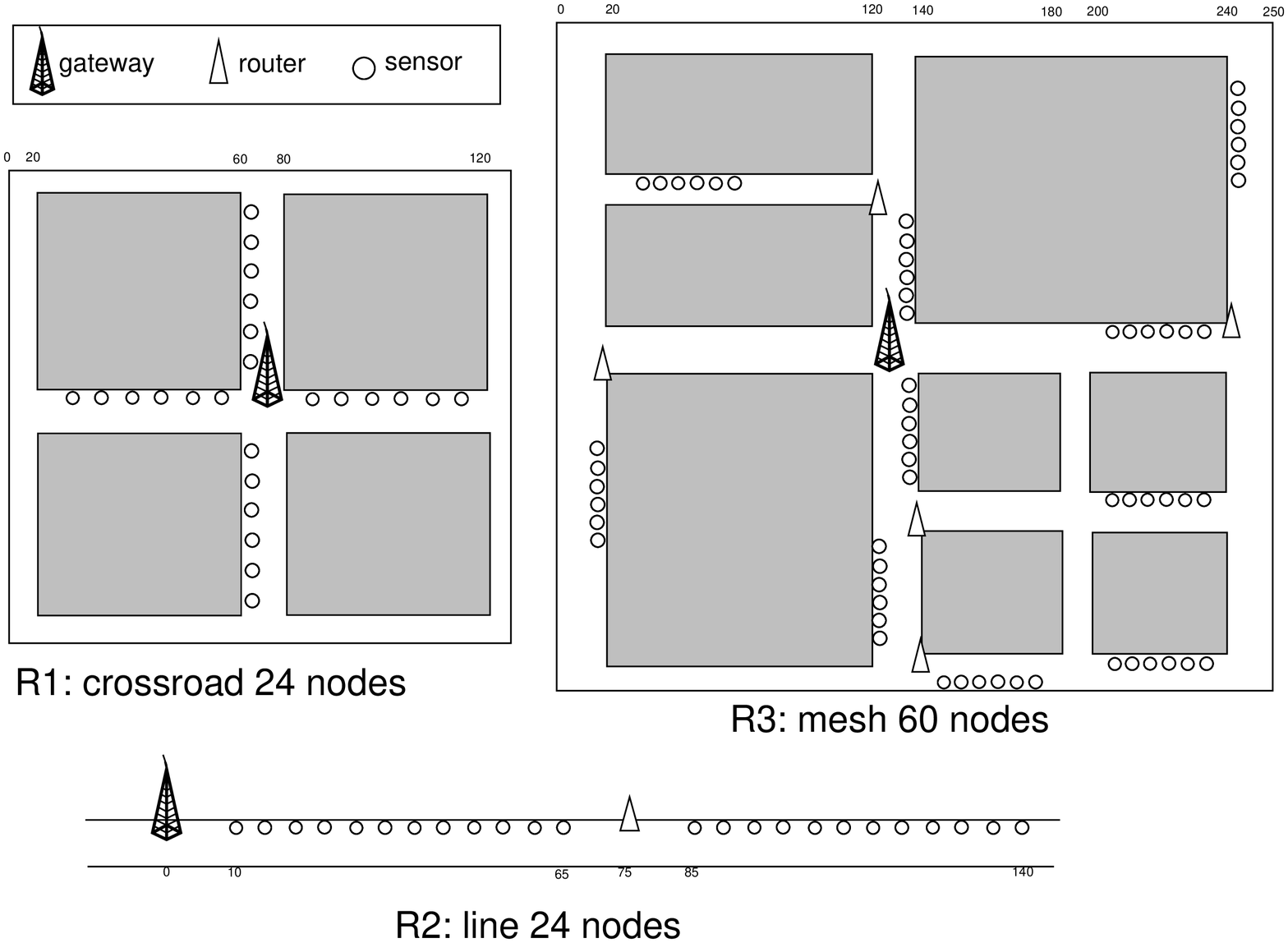}
\begin{tabular}{l}
R1 -- Crossroad: 1 gateway \& 24 parking sensors  \\
R2 -- Line: 1 gateway, 1 router \& 24 parking sensors  \\
R3 -- Mesh: 1 gateway, 5 router \& 60 parking sensors \\
\end{tabular}
\caption{Three different topologies for our simulations}
\label{fig:topology-region.eps}
\end{figure}

From the real implementation in SFpark project \cite{SFparkreport} and in SmartSantander \cite{SmartSantanderReport}, the network deployment is quite consistent. All the wireless parking sensors are in-ground and have limited communication among themselves because vehicles are obstacles for wireless communication. Thus, sensors regarded as RFD (reduced function device) can only communicate with routers or gateways which are FFD (full function device) and generally mounted on the streetlights or traffic lamps. We configured 3 different topologies by referring to the three regions R1--R3 in \cite{SmartSantanderReport} with traffic parameters. The impact of the traffic intensity is in our previous report \cite{lin:hal-00948120}. To clarify the influence of four different MAC protocols, we choose one set of parameters from them for our scenarios. The topologies are depicted in figure~\ref{fig:topology-region.eps}. Our simulations, performed with the WSNet Simulator \cite{WSNet}, use the simulation parameters in table~\ref{tab:SimulationParameters}.

\begin{table}[!t]
\caption{Simulation parameters}
\label{tab:SimulationParameters}
\small
\centering
	\begin{tabular}{|l|l|l|l|}
	\hline
	\multicolumn{2}{|l|}{Simulation time: 10000 seconds} & \multicolumn{2}{l|}{Experiences: \#20}\\	
	\hline
	\multicolumn{4}{|l|}{Distance between two adjacent sensors: 5 meters} \\
	\hline
	\multicolumn{2}{|l|}{Transmit power output 3 dBm} & \multicolumn{2}{l|}{Receive sensitivity -85 dBm} \\
	\hline
	\multicolumn{2}{|l|}{Data rate 250 kbps} & \multicolumn{2}{l|}{802.15.4 Radio} \\
	\hline
	$P_{tx}$ 65.7 mW  & $P_{rx}$ 56.5 mW & $P_{cs}$ 55.8 mW& $P_{off}$ 30 $\mu$W \\
  \hline
  \multicolumn{2}{|l|}{$E_{radio.switch}$ 0.16425mJ} & \multicolumn{2}{l|}{Packet size 84 bytes} \\
	\hline      
  \multicolumn{4}{|l|}{MAC: duty-cycled $s_D = 0.1s$. Retransmission \& piggyback.} \\
	\hline
	\multicolumn{4}{|l|}{Routing: gradient \cite{Watteyne-5425543}} \\
	\hline
	\multicolumn{4}{|l|}{Pathloss: Corner propagation ($\lambda = 0.125$) \cite{1492678} + Rayleigh fading} \\
	\hline
	\multicolumn{4}{|l|}{Vehicle event parameters: } \\
	\multicolumn{4}{|l|}{$T_p$ parking time $\sim$ Weibull($x_1$; $\alpha_1$, $\lambda_1$). $\alpha_1 = 0.4$. $\lambda_1 = 60*60$.} \\
	\multicolumn{4}{|l|}{$T_v$ vacant time  $\sim$ Weibull($x_2$; $\alpha_2$, $\lambda_2$). $\alpha_2 = 0.7$. $\lambda_2 = 15*60$.} \\
	\hline
	\multicolumn{4}{|l|}{Hybrid event-driven and periodic application: $\omega = 120s$. $\tau=30s$} \\
  \hline
	\end{tabular}
\end{table}

\subsection{Packet interarrival time}
 
\begin{figure}[!t]
\begin{minipage}{\linewidth}
\centering
\includegraphics[height=\linewidth, angle=-90]{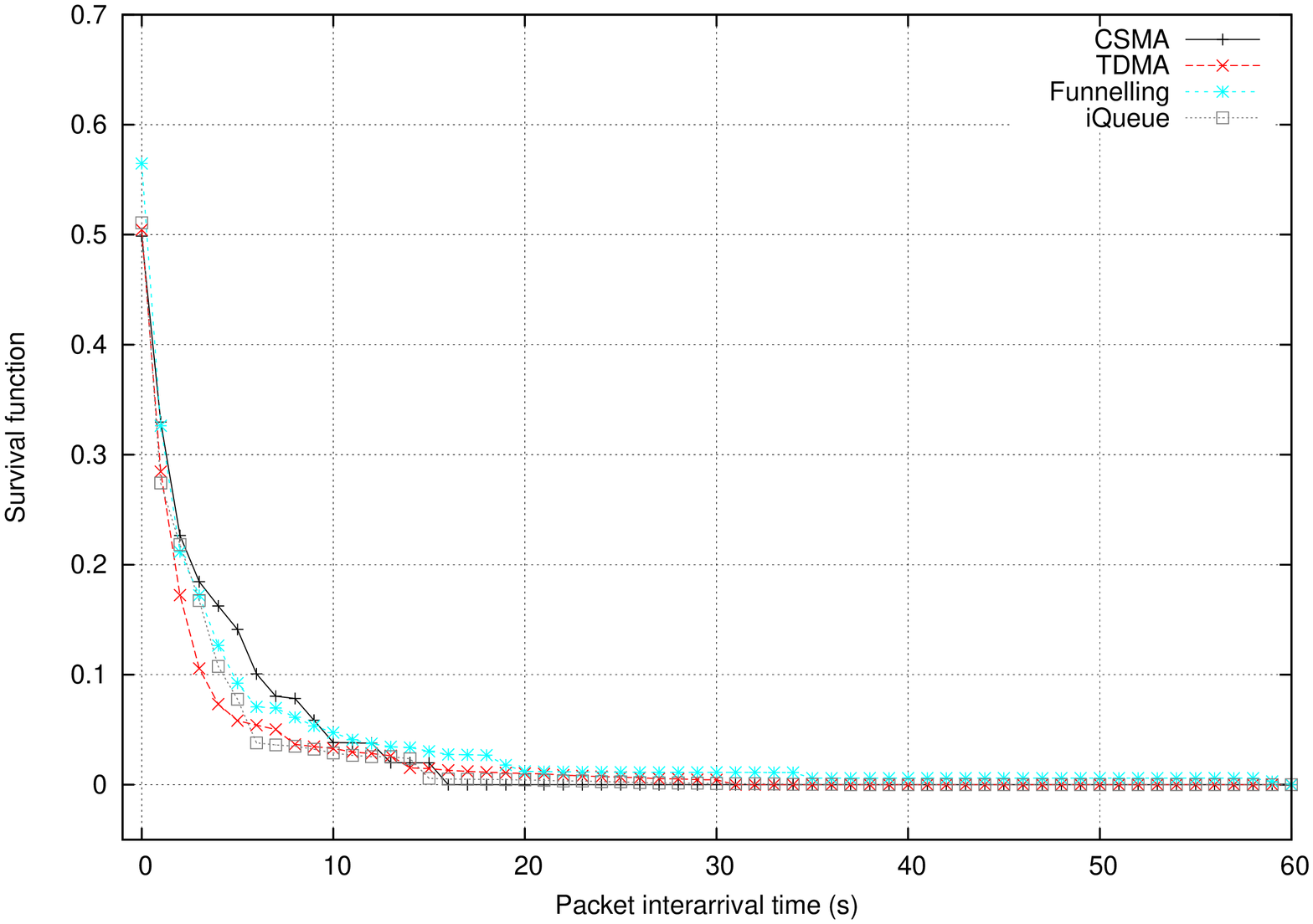}
\caption{The survival function of packet interarrival time on the gateway side: Crossroad}
\label{fig:crossroad-packet-interarrival-time}
\end{minipage}
\begin{minipage}{\linewidth}
\centering
\includegraphics[height=\linewidth, angle=-90]{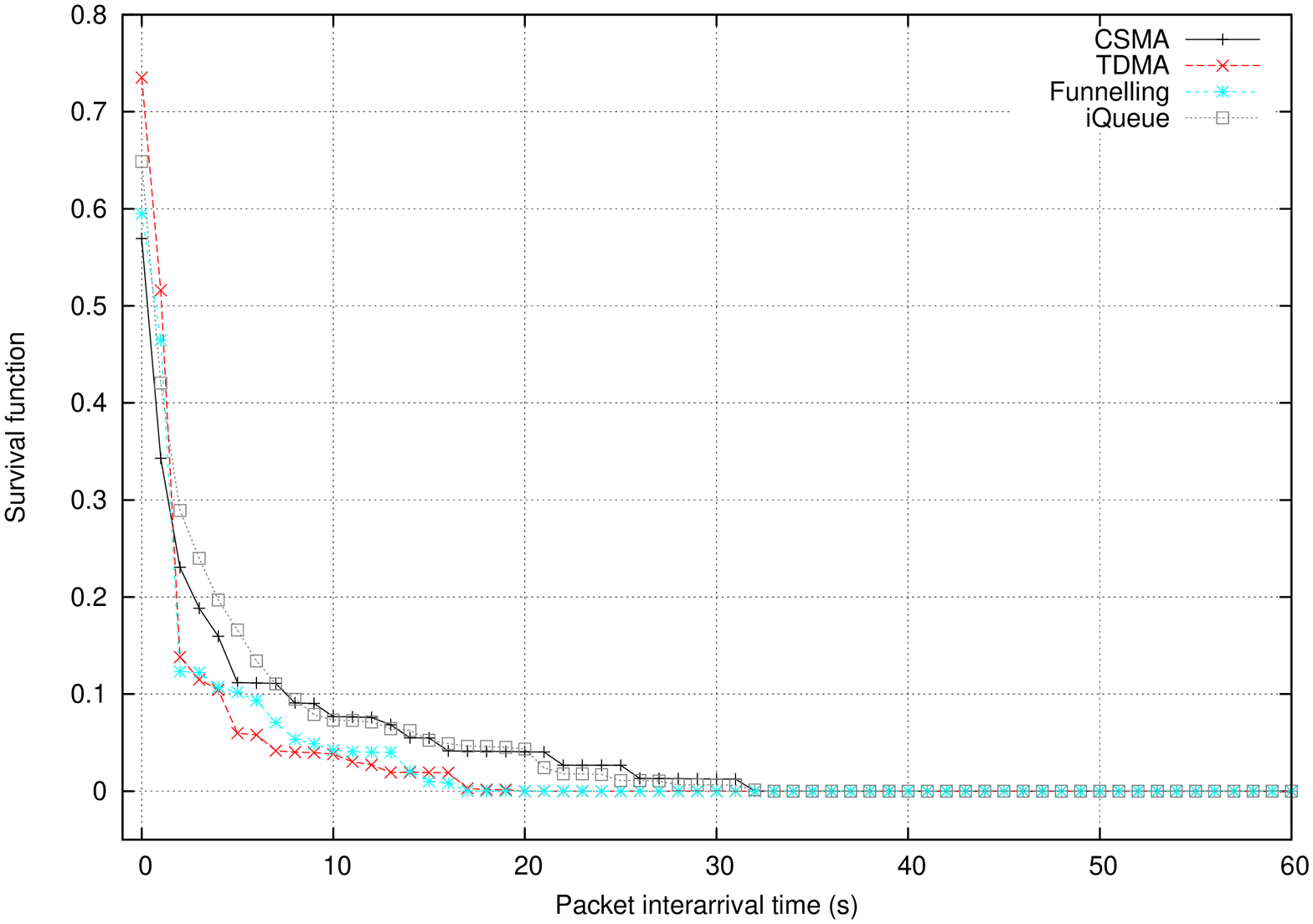}
\caption{The survival function of packet interarrival time on the gateway side: Line}
\label{fig:line-packet-interarrival-time}
\end{minipage}
\begin{minipage}{\linewidth}
\centering
\includegraphics[height=\linewidth, angle=-90]{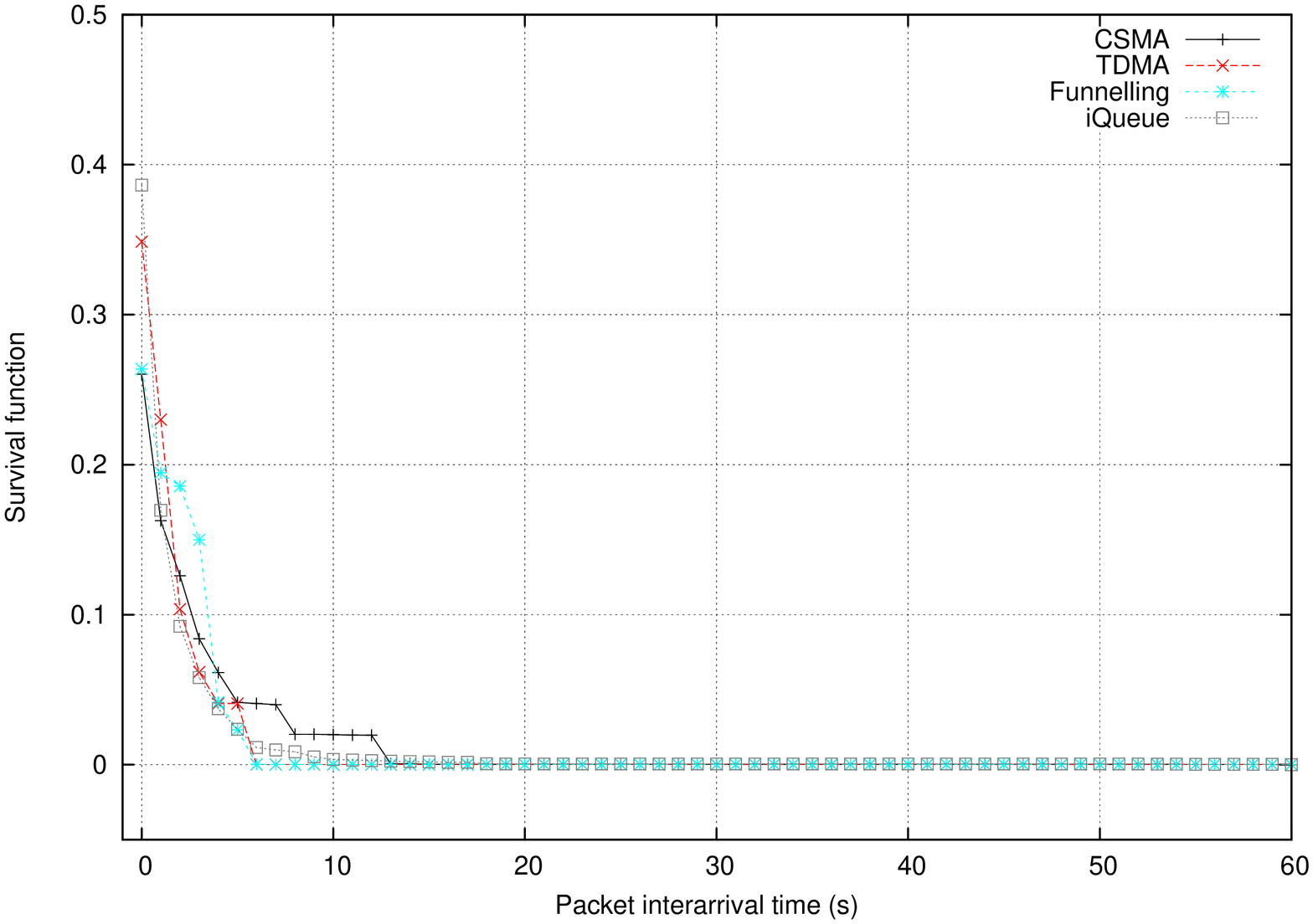}
\caption{The survival function of packet interarrival time on the gateway side: Mesh}
\label{fig:mesh-packet-interarrival}
\end{minipage}
\end{figure}

Figures~\ref{fig:crossroad-packet-interarrival-time} and~\ref{fig:line-packet-interarrival-time} show the packet interarrival time of 24 wireless parking sensors in two different topologies, namely crossroad and line types. Compared with figure~\ref{fig:packet-vehicle-interarrival-time-24-nodes}, the interarrival time is reformed and the average value is until 30 according to the traffic threshold $\tau$ which decides when an updated information is delivered. The contention-based protocols restore better the curves to the original form which is the exponential distribution, thanks to its traffic adaptive property. In the topology R2, parts of parking sensors are far away from the gateway and can only reach the gateway by the router in the middle of the street through two-hop. This way, we see that CSMA and i-Queue MAC both give similar interarrival time which is close to the original form in figure~\ref{fig:line-packet-interarrival-time}. On the contrary, TDMA and Funnelling, applied schedule-based around the gateway, have a longer delay time and differ from the other two. Also some packet interarrival time is over 30 seconds because of the longer transmission distance. Figure~\ref{fig:mesh-packet-interarrival} can be regarded as a mix-up of crossroad and line types. As the network size increases, so does the probability of packet arrival. It is obvious that the different MAC protocols cause a slight difference, however, the interarrival time is still approximately to exponential distribution. Hence, the Markov's queueing model may be possible to be applied.

\begin{figure}[!t]
\begin{minipage}{\linewidth}
\centering
\includegraphics[height=\linewidth, angle=-90]{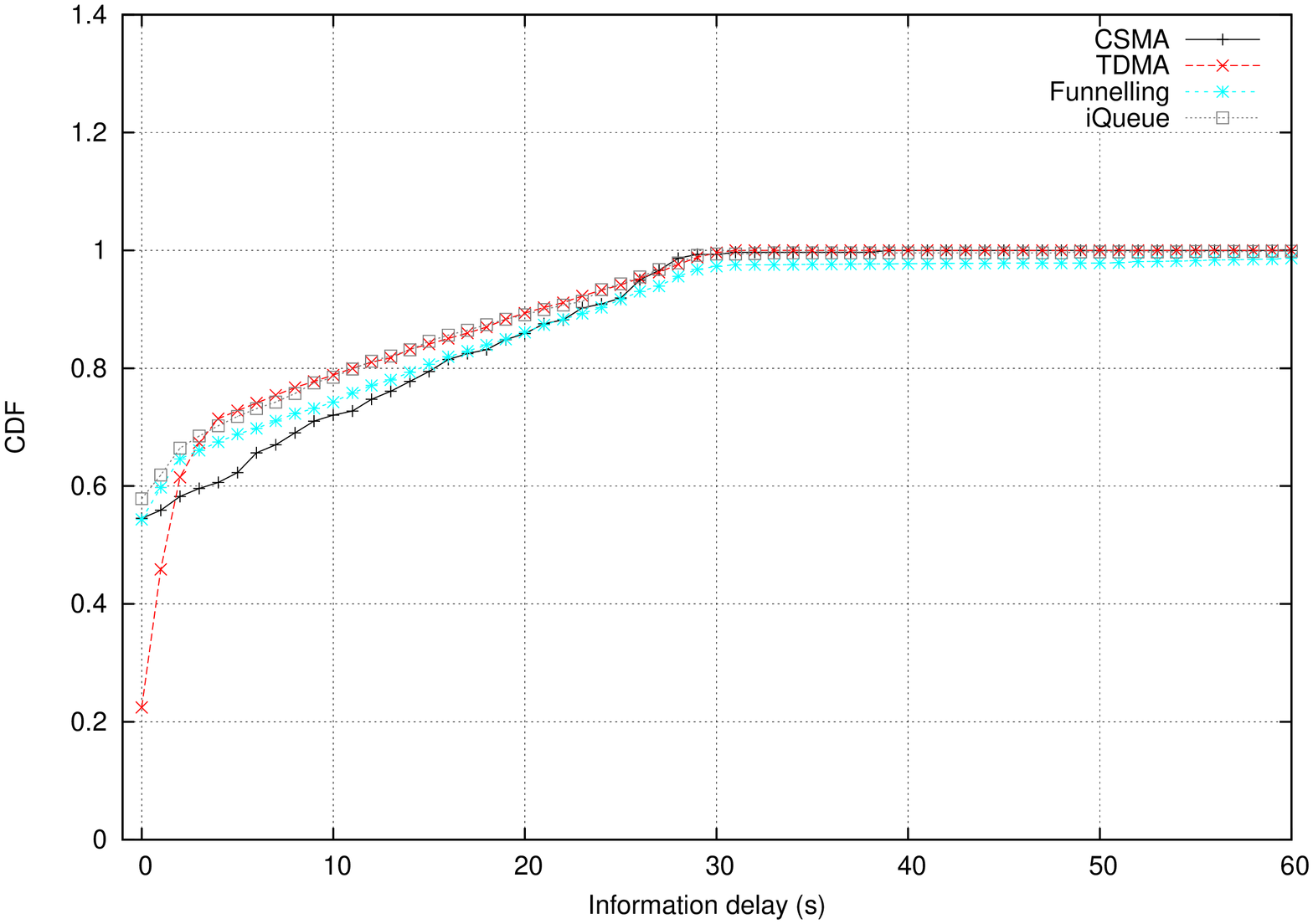}
\caption{The information delay of changed parking occupancy status: Crossroad}
\label{fig:crossroad-info-delay-time}
\end{minipage}
\begin{minipage}{\linewidth}
\centering
\includegraphics[height=\linewidth, angle=-90]{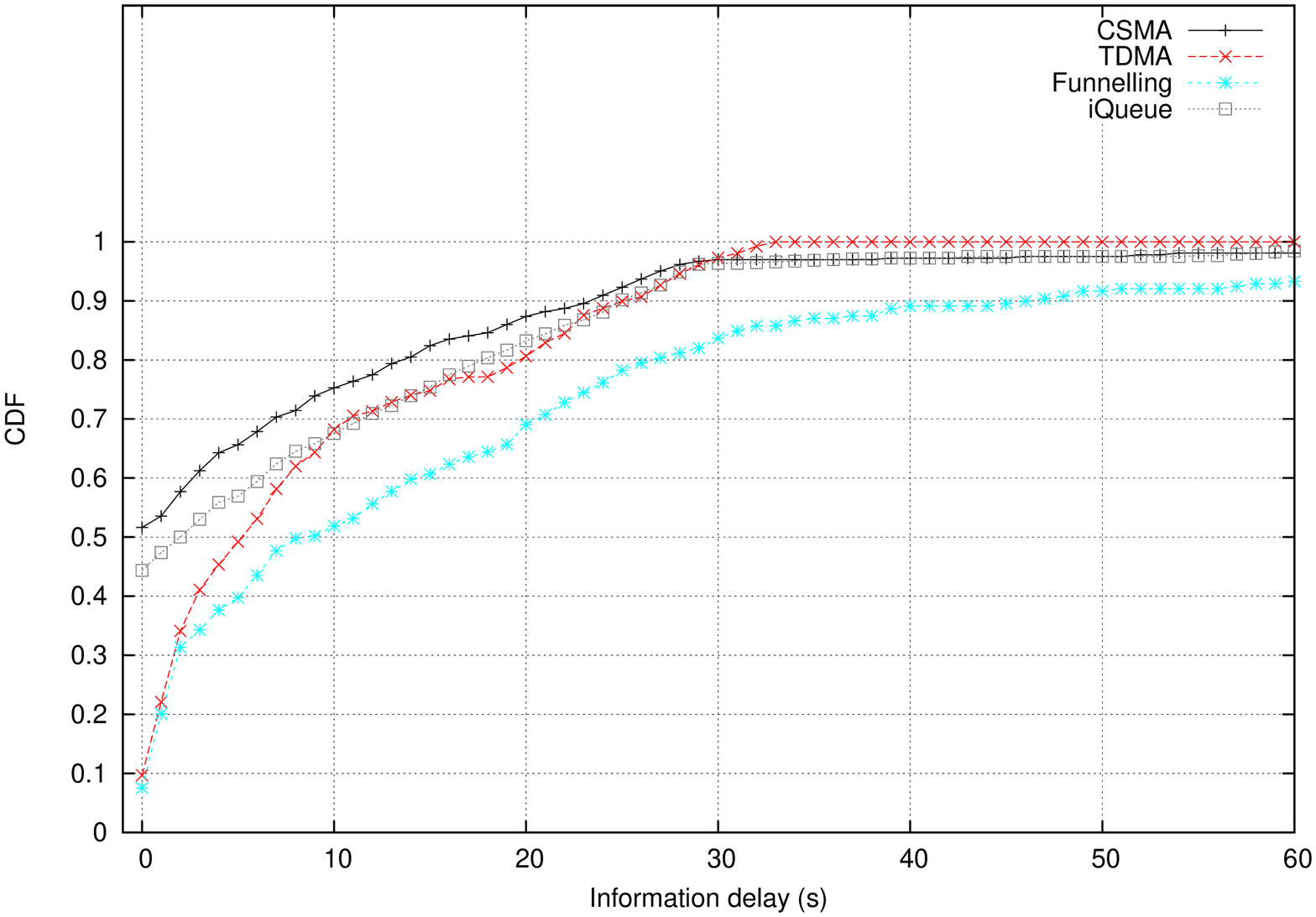}
\caption{The information delay of changed parking occupancy status: Line}
\label{fig:line-info-delay-time}
\end{minipage}
\begin{minipage}{\linewidth}
\centering
\includegraphics[height=\linewidth, angle=-90]{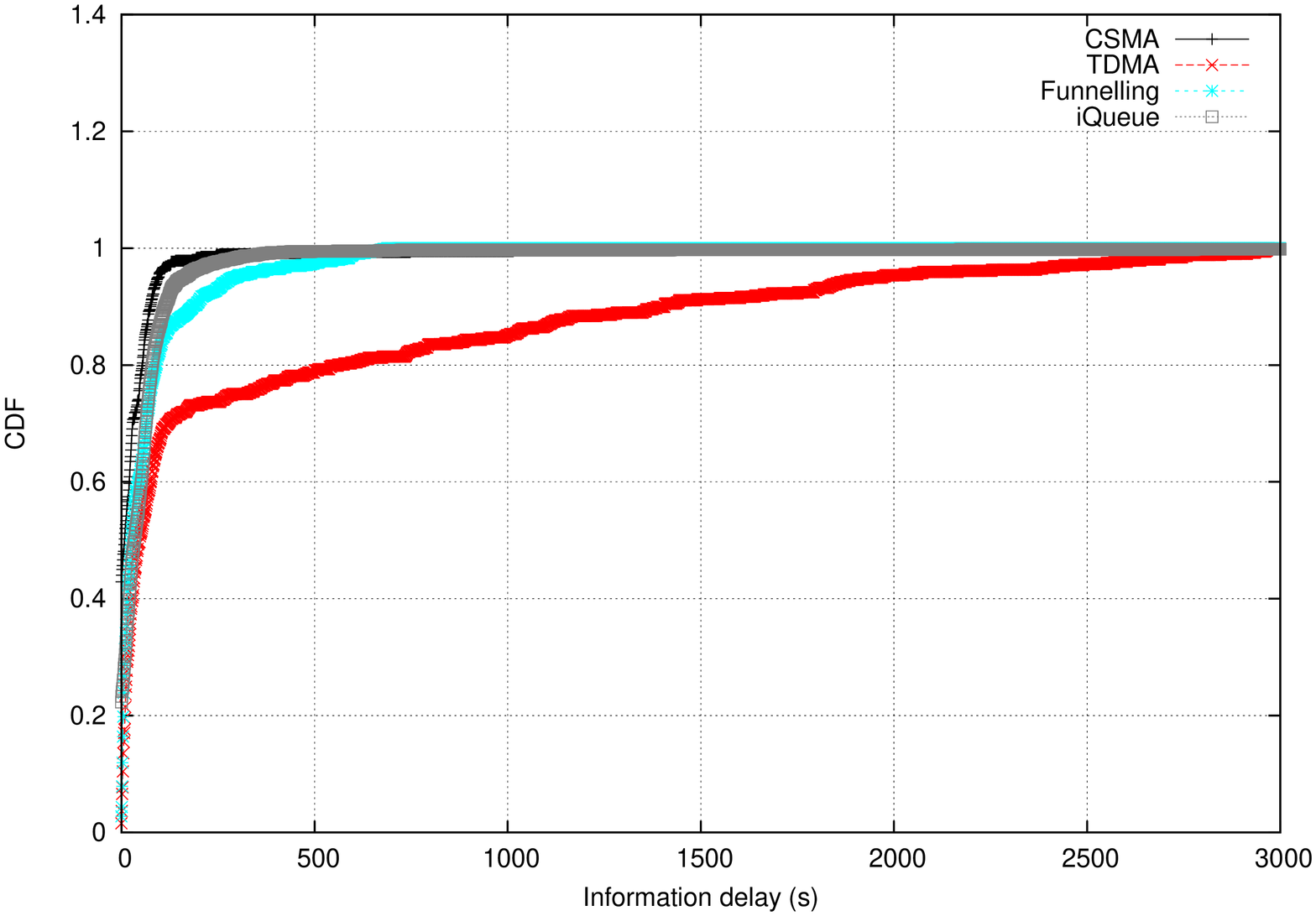}
\caption{The information delay of changed parking occupancy status: Mesh}
\label{fig:mesh-info-delay}
\end{minipage}
\end{figure}

\subsection{Information delay}

Smart parking service provides real-time information to drivers thanks to the unceasing parking sensing. Thus, the change of parking occupancy status must arrive at the gateway at a shortest delay. In the topology of R1, each parking sensor can reach the gateway in one hop. In figure~\ref{fig:periodic-trigger-mixed-app} we showed the two turning points of periodic and event-driven hybrid application, namely at the time point of one-time duty-cycle and switching threshold $\tau$. The information delay of the topology R1 is shown in figure~\ref{fig:crossroad-info-delay-time}, and only TDMA shows clearly the two turning points since its duty-cycle period is longer than the others. While changing to the linear topology R2, TDMA has almost no difference but prolongs its information delay by one-hop more communication. Contention-based protocol basically performs a better delay time in the beginning but then less good that TDMA as time goes. Funnelling-MAC assigns the nodes one-hop around a guaranteed time slot uniformly but it does not consider the different traffic loads among them. In R2, the router and all the parking sensors, between the gateway and the router, are all assigned their exclusive time slot once. However, the router, who has to forward all the generated packets in its right side, will have much more packets to send than its neighbors. In a word, there is a bottleneck on the router and it drops the network delay time distinctly. i-Queue MAC basically adapts to the network traffic according to the queue indicator. But restricted to the vTDMA period, the contention access period is shortened and it provokes some more packet collisions. Normally, the vTDMA period is expected to help to reduce the CSMA-slot competitors, but it is less common that a node has two consecutive packets with the sporadic traffic in PSNs except the routers. Thus sensors compete in a shorter contention access period and packet collision happens more frequently. Then if we increase the network dimension, it implies that there are more routers around the gateways. The result in figure~\ref{fig:mesh-info-delay} shows that i-Queue MAC has a good performance as CSMA but the slot allocation still requires at least one packet transmission in the CSMA time slots of CAP. TDMA has a very long information delay because of its longer duty-cycle and also the construction of multiple-hop scheduling. The schedule can be optimized when the network topology is known. 

\subsection{Lifetime}

\begin{figure}[!t]
\begin{minipage}{\linewidth}
\centering
\includegraphics[height=\linewidth, angle=-90]{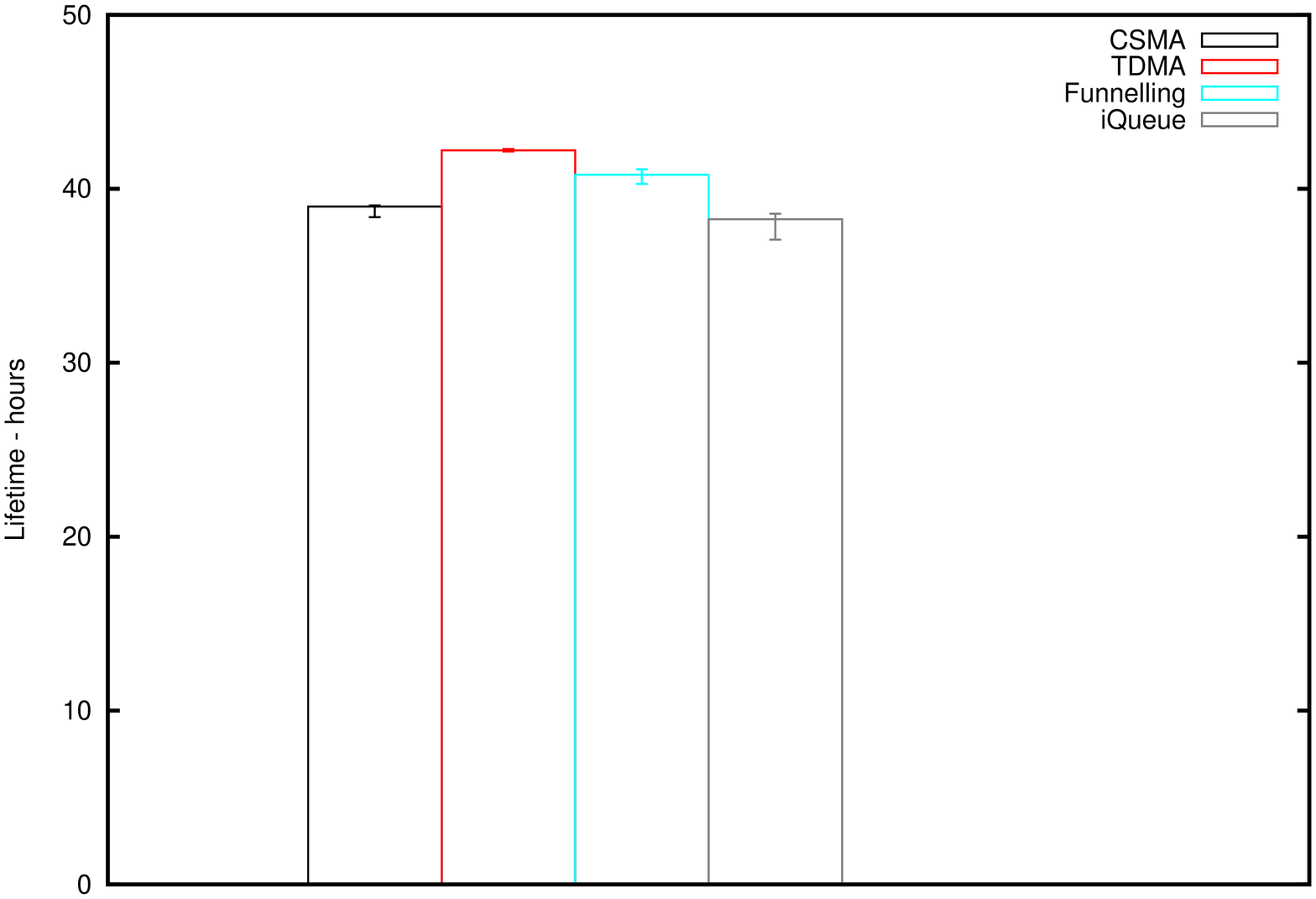}
\caption{The lifetime of parking sensors while $s_D = 0.1s$ and battery capacity 6.3Ah: Crossroad}
\label{fig:crossroad-energy-report}
\end{minipage}
\begin{minipage}{\linewidth}
\centering
\includegraphics[height=\linewidth, angle=-90]{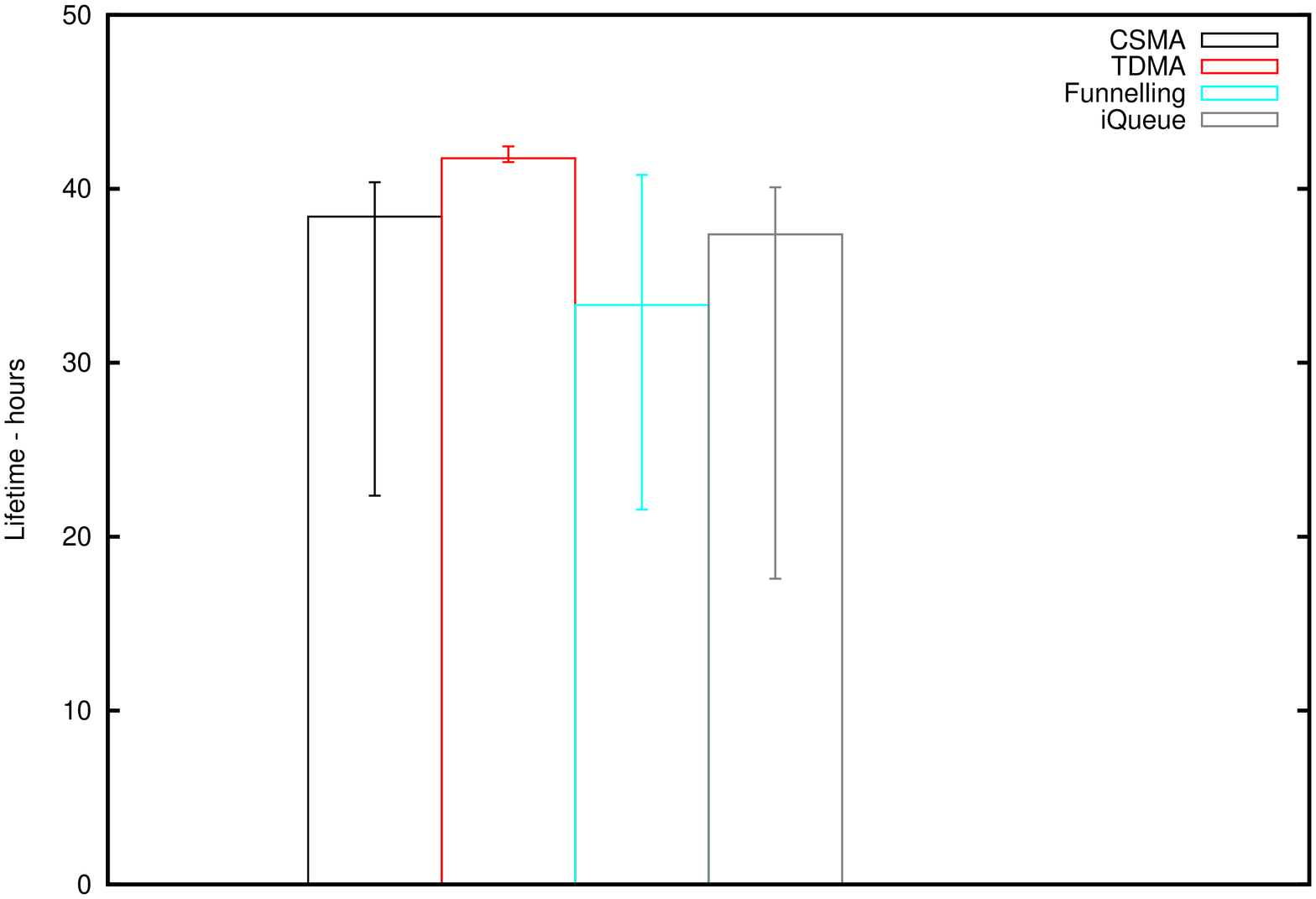}
\caption{The lifetime of parking sensors while $s_D = 0.1s$ and battery capacity 6.3Ah: Line}
\label{fig:line-energy-report}
\end{minipage}
\begin{minipage}{\linewidth}
\centering
\includegraphics[height=\linewidth, angle=-90]{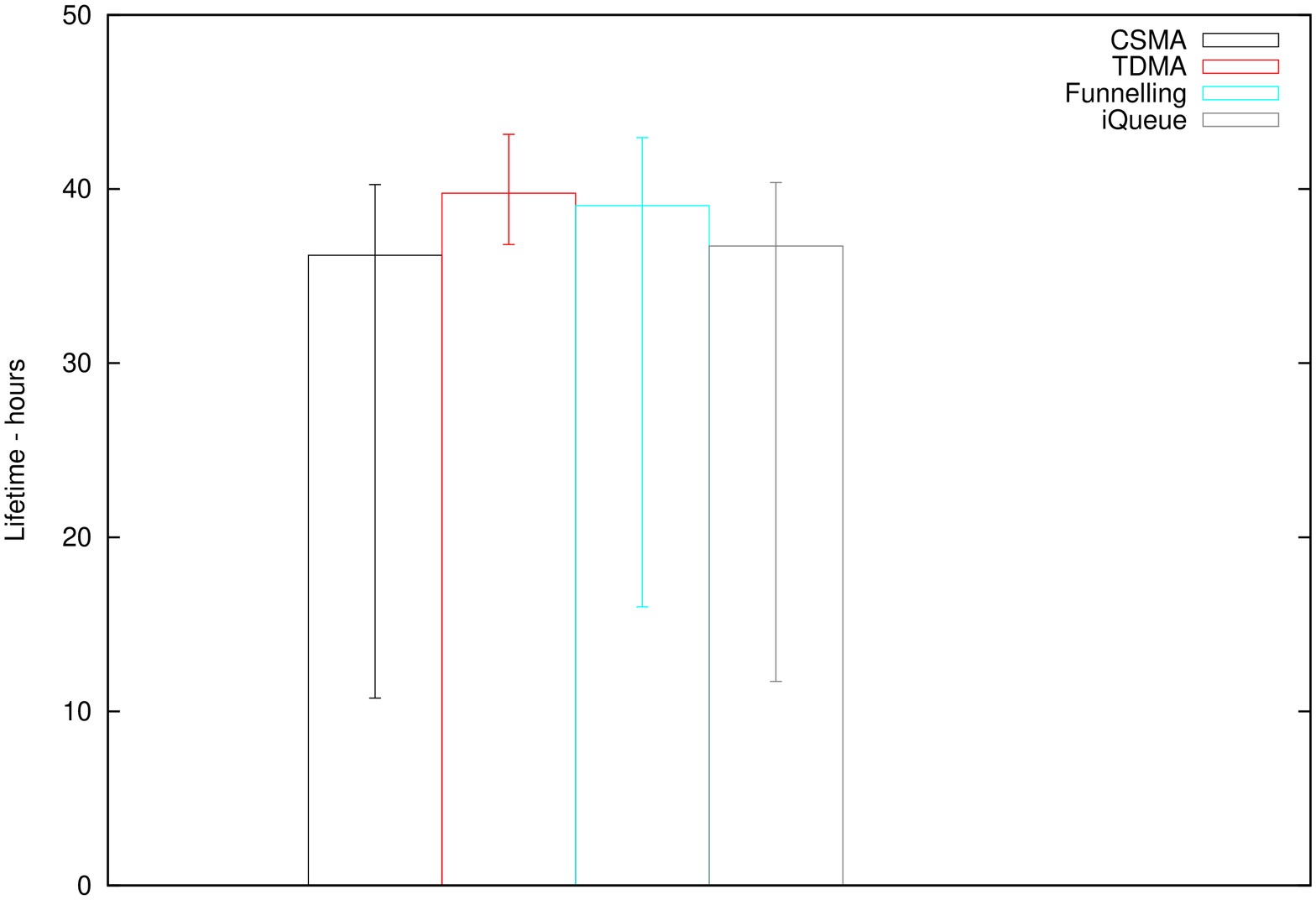}
\caption{The lifetime of parking sensors while $s_D = 0.1s$ and battery capacity 6.3Ah: Mesh}
\label{fig:mesh-energy-report}
\end{minipage}
\end{figure}

Finally, we look at the autonomy of parking sensor network. The most important is the energy depletion. If the parking sensor cannot live for a long time, the maintenance of the infrastructure will be expensive for the city. We limit the battery capacity to 6300mAh, the slot duration 0.1 seconds and $T_{GTS}+T_{inactive} = 0$, then compare the lifetime of parking sensors under different bandwidth allocation strategies. Figure~\ref{fig:crossroad-energy-report} shows the lifetime in the topology R1 crossroad. Since each parking sensor can reach the gateway through one hop, the packet collision rate is low and so does the energy deviation. Funnelling-MAC is between TDMA and CSMA according to the percentage of TDMA slots ($n_{tdma}$) and CSMA slots ($n_{csma}$). Here we added Rayleigh fading in the propagation model and some nodes can probably run in CSMA mode if they do not receive the slot allocation messages in the beginning. i-Queue MAC behaves like CSMA but consumes a bit more energy as a result of few contention time slots and additional size in the i-Queue MAC header. 

Figure~\ref{fig:line-energy-report} shows the energy consumption of the linear topology R2. It is obvious that all MAC protocols have a higher energy depletion deviation except TDMA. That is because TDMA is the only contention-free (schedule-based) MAC protocol in our scenarios and half of nodes have to run in multi-hop network which provokes an hidden terminal problem. Funnelling-MAC has a very bad performance in linear topology due to the unbalanced TDMA slot allocation for the router. 

Figure~\ref{fig:mesh-energy-report} shows the lifetime of parking sensor in mesh topology. The energy depletion variation of TDMA varies more than the other two topologies because the larger network dimension costs more energy during the signaling period. i-Queue MAC benefits from more routers around and reduces the competitors in the network. However, its energy variation is still as much as CSMA.  Funnelling-MAC has a quite good energy efficiency in the mesh network if each linear path is not too long. Hence, to improve the Funnelling-MAC, the routing protocol will also be concerned.

\section{Conclusions}
\label{sec:conclusion}

In this paper, we studied the wireless parking sensor networks from the viewpoint of network traffic and MAC protocols. The traffic model is generally affected by vehicle's arrival and departure which are both heavy-tailed. While looking at a group of parking sensors, the simulation shows that the packet interarrival time is no longer heavy-tailed but exponentially distributed, which might help the reuse of Markov's queueing model. Then we chose four different types of MAC protocols and evaluated them in 3 kinds of network topologies. The result show that an hybrid adaptive MAC can help to improve the network performance by providing a better energy-delay tradeoff. However, it can also drop the network performance when the topology changes. Funnelling-MAC is a quite good choice for the crossroad and mesh topologies, but not for a long linear one. The other hybrid MAC protocol, i-Queue MAC, only works well in small cell and one-hop network because the burstiness of PSN is caused by increasing transmitters, not the traffic load of one sensor node. The important criteria of choosing an appropriate MAC protocol are to consider the deployment of sensors and routers which connect to gateway directly and their traffic models, like the packet interarrival time can presents the bursty network traffic on different groups of nodes. From the result, it also shows that the information delay is bounded by application and MAC parameters, even with the bursty traffic generated by Weibull distribution. 

\section{Acknowledgments}
Funding for this project was provided by two grants from the Rh\^one-Alpes Region, France: F. Le Mou\"el currently holds a mobility grant Explora'Pro and T. Lin a doctoral fellowship ARC 07 n$^{o}$7075.

\raggedright

{\small
\bibliography{sigproc}  
\bibliographystyle{abbrv}
}


\end{document}